\documentclass[superscriptaddress,aps,pra,twocolumn,preprintnumbers,amsmath,amssymb,floatfix,groupedaddress]{revtex4}

\usepackage{enumerate}
\usepackage{graphicx}
\usepackage{color}

\newcommand{\beq}{\begin{equation}}
\newcommand{\eeq}{\end{equation}}
\newcommand{\bea}{\begin{eqnarray}}
\newcommand{\eea}{\end{eqnarray}}
\newcommand{\ba}{\begin{array}}
\newcommand{\ea}{\end{array}}

\newcommand{\bef}{\begin{figure}}
\newcommand{\eef}{\end{figure}}

\begin{document}

\title{Recovering the quantum formalism from physically realist axioms. }
\author{Alexia Auff\`eves$^{(1)}$ and Philippe Grangier$^{(2)}$} 
\affiliation{
(1): Institut N\' eel$,\;$BP 166$,\;$25 rue des Martyrs$,\;$F38042 Grenoble Cedex 9, France. \\
(2): Laboratoire Charles Fabry, IOGS, CNRS, Universit\'e Paris~Saclay, F91127 Palaiseau, France.
 }

%\affil[+]{these authors contributed equally to this work}

%\keywords{Keyword1, Keyword2, Keyword3}

\begin{abstract}
We present a heuristic derivation of Born's rule and unitary transforms in Quantum Mechanics, from a simple set of axioms built  upon a physical phenomenology of quantization. This approach naturally leads to the usual quantum formalism, within a new realistic conceptual framework that is discussed  in details. Physically, the structure of Quantum Mechanics appears as a result of the interplay between the quantized number of  ``modalities"  accessible to a quantum system, and the continuum of  ``contexts" that are required to define these modalities. Mathematically, the Hilbert space structure appears as a consequence of a specific  ``extra-contextuality" of modalities,  closely related to the hypothesis of Gleason's theorem, and consistent with its conclusions.
\end{abstract}

\maketitle

\section*{Introduction}

A series of recent experimental tests of Bell's theorem \cite{B1,B2,B3}  have been said to close the door on Einstein's and Bohr quantum debate \cite{AA}. It is generally considered that Einstein lost the case, by advocating a notion of ``local realism" incompatible with quantum mechanics (QM) \cite{EPR}. However, Bohr \cite{NB} also presented himself as a realist as far as physics is concerned, and QM has no direct conflict with relativistic causality. One may thus wonder whether a deep -- but philosophically sound --  redefinition of physical reality might provide a way  to reconcile the founding fathers of quantum physics. In refs.\cite{csm1,csm3,ph1,ph2} we argued that this can be done, under the condition that fully predictable physical properties (called ``elements of physical reality" in ref.\cite{EPR}) are attached not to a system alone, but to a system within a given experimental context \cite{NB}.

In this paper, we further exploit this idea, in order to present a heuristic derivation of the quantum formalism, understood as a non-classical way to calculate probabilities. 
An outstanding  feature of our approach is that the superposition principle  and  Born's rule 
appear as consequences of the quantized number of states accessible to a quantum system,
without any appeal to ``wave functions" -- but we do recover projective measurements. 
%For our purpose we will concentrate on  
%pure states and projective (orthogonal) measurements. 
Our approach bears some relationship with Gleason's theorem 
\cite{gleason,helena,csm2}, as it will be discussed below (see also Methods).

We shall start without formalism, but from a few definitions and hypotheses, 
presented here as axioms. These axioms are based on standard quantum phenomenology, 
and they have been introduced and discussed  in  \cite{csm1,csm3}
under the acronym ``CSM", meaning Context, System, Modality. 
In the present article we will not repeat this discussion, but rather use the the following axioms
to summarize the  main features of our approach.
Though the formulation of the axioms contains very little mathematics, they have deep mathematical 
consequences, that will be spelled out in the ``Results" section below. 

%have been discussed in ref.\cite{csm1}, both physically and philosophically. 

\begin{itemize}

\item {\bf Axiom 1} (modalities): 
(i) Given a physical system, a {\it modality} is defined as the values of a complete set of physical quantities  that can be predicted with certainty and measured repeatedly on this system. 

(ii) Here a ``complete set" means the largest possible set compatible with certainty and repeatability, for all possible modalities attached to this set. This complete set of physical quantities  is called a {\it context}, and the  modality  is attributed {\bf jointly}  to the system and the context. 
(iii) Modalities cannot show up independently of a context, but 
the same modality may appear in different contexts, with the same conditions of repeatability and certainty.

\item {\bf Axiom 2} (quantization):  (i) For a given context, 
there exist $N$ distinguishable modalities $\{u_i\}$, that are mutually exclusive~: if one modality is true, or realized, 
the others are wrong, or not realized. (ii) The value of $N$, called the dimension, is a characteristic 
property of a given quantum system, and is the same in all relevant contexts.

\item {\bf Axiom 3} (changing contexts):  Given axioms 1 and 2, 
the different contexts relative to a given quantum system are related between themselves 
by continuous  transformations $g$ which 
are associative, have a neutral element (no change), and an inverse.
Therefore the set of context transformations $g$ has the structure of a continuous group ${\cal G}$.

\end{itemize}

For the sake of clarity, we  note
that, within the usual QM formalism (not used so far), a modality and a context correspond
respectively to a pure quantum state, and to a complete set of commuting observables. 
The axioms are formulated for a finite $N$, but this restriction will be lifted below.  
Intuitively, as discussed in details in  ref.\cite{csm1},   
a context can be seen as a given ``knob settings" of the measurement apparatus. 
%These axioms, under the acronym ``CSM", meaning Context, System, Modality,
%have been discussed in ref.\cite{csm1}, both physically and philosophically. 
We will not repeat this discussion here, but we want to consider  the following question:
%Here we want to draw consequences from them, by considering  the following question:
it is  postulated in Axiom 2  that there are $N$ mutually exclusive modalities 
associated to each given context,   but there are many more modalities, 
corresponding to all possible contexts, related according to Axiom 3. These modalities are generally not mutually exclusive, 
but are {\bf incompatible}: it means that if one is true, one cannot tell whether the other one  is true or wrong.
Then, how to relate between themselves all these modalities ? 
 
A first crucial result already established in ref.\cite{csm1} is that this connection can only be a probabilistic one,  otherwise the axioms would be  violated; the argument is as follows.
Let us consider a single system, two different contexts $C_u$  and $C_v$, and the associated modalities $u_i$ and $v_j$, where $i$ and $j$ go from 1 to $N$. The quantization principle (Axiom 2) forbids to gather all the modalities $u_i$ and $v_j$ in a single set of more than $N$ mutually exclusive modalities, since their number is fixed to $N$. Therefore the only relevant question to be answered by the theory is: If the initial modality is $u_i$ in context $C_u$, what is the {\it conditional probability} for obtaining modality $v_j$ when the context is changed from $C_u$ to $C_v$ ? We emphasize that this probabilistic description is the unavoidable consequence of the impossibility to define a unique context making all modalities mutually exclusive, as it would be done in classical physics. It appears therefore as a joint consequence of the above Axioms 1 and 2, i.e. that modalities are quantized, and require a context to be defined. 

Now, according to Axiom 3, changing the context results from changing the measurement apparatus 
at the macroscopic level, that is, ``turning knobs". A typical example is changing
the orientations of a Stern-Gerlach magnet. These context transformations have the mathematical structure of a continuous
group, denoted ${\cal G}$~: the combination of several transformations is associative  and gives a new transformation,
there is a neutral element (the identity), and each transformation has an inverse. 
Generally this group is not commutative : for instance, the three-dimensional
rotations associated with the orientations of a Stern-Gerlach magnet do not commute.
For a given context, there is a given set of $N$ mutually exclusive modalities, denoted $\{u_i\}$. 
By changing the context,  one obtains $N$ other mutually exclusive modalities, denoted $\{v_j\}$, 
and one needs to build up a mathematical formalism, able  to provide  
the probability that a given initial modality $u_i $ ends up in a new modality $v_j $.

The standard approach at this point is to postulate that each modality $u_i$  is associated with a
vector $|u_i \rangle$  in a $N$-dimensional Hilbert space, and that the set of $N$  mutually 
exclusive modalities in a given context  is associated to a set of $N$ orthonormal vectors.
Rather than  vectors $|u_i \rangle$ and $|v_j \rangle$,  one can 
equivalently use rank-one projectors  $ P_{u_i } $ and $P_{ v_j  }$, and 
Born's rule giving the conditional probability $p(v_j | u_i)$ can be written as  
\beq
p(v_j | u_i) = \mathbf{Trace}( P_{ u_i } P_{ v_j }).
\label{born1}
\eeq

In this article, we will postulate neither Born's rule nor even Hilbert spaces, 
but  we will derive them as the consequence of the previous Axioms. 
Then we  will discuss the relation with Gleason's theorem, 
as well as  some consequences of our approach. 

\section*{Results}
%\vskip 2mm
In this  part, we start from Axioms 1-3 and construct a consistent probability theory,
by imposing some requirements on what it should describe. 
The first steps will thus be to translate the Axioms into 
mathematical constraints on probabilities relating  modalities. 
This will lead us  to manipulate $N \times N$ probability matrices, in a general way not restricted to the quantum formalism.
Using our Axioms to obtain physically-based constraints, we will finally get
Born's rule and unitary transforms.  
\\

%\subsection*{The general probability matrix.}
\noindent \textbf{The general probability matrix.}
\vskip 2mm

Let us denote $\{u_i\}$ and $\{v_j\}$ the respective modalities of the initial and final context, %}
and define the probability $p_{v_j|u_i} $ of finding the particular modality $v_j$,
when starting  from modality $u_i$.  There are $N^2$ such probabilities, 
that  can be arranged  in a $N \times N$ matrix $\Pi_{v|u} = \left( p_{v_j|u_i} \right)$, containing
 all  probabilities  connecting the $N$ modalities in each context  $\{u_i\}$ and $\{v_j\}$. 
Since one has obviously $0 \leq p_{v_j|u_i}  \leq 1$ and 
$\Sigma_j \; p_{v_j|u_i} = 1$, the matrix $\Pi_{v|u}$ is said to be a {\it stochastic} matrix
(see Methods for definitions). 

For clarity, let us emphasize  the interpretation of the conditional probability notation:
in agreement with the definition of modalities as certainties, 
 the meaning of $p_{v_j|u_i}$ is  that ``if we start  (with certainty) from modality $u_i$
 in the old context, then the probability to get modality $v_j$ in the new context is $p_{v_j|u_i}$". 
These probabilities provide the connection between
theoretical predictions  and experiments, and correspond 
to relative frequencies in repeated experiments starting from 
 the same  $u_i$. It  is not critical whether they are  interpreted in a frequentist or Bayesian sense, but 
it is critical to acknowledge that  they are intrinsic consequences of our axioms on quantized modalities, 
and thus are {\bf not} associated to any ``missing information". 

\noindent For $N=3$, one has for instance
$$\Pi_{v|u} = \left(  \begin{array}{lll} 
 p_{v_1|u_1}, & p_{v_2|u_1}, & p_{v_3|u_1}  = 1 -  p_{v_1|u_1}- p_{v_2|u_1}\\ 
 p_{v_1|u_2}, & p_{v_2|u_2}, & p_{v_3|u_2}  = 1 -  p_{v_1|u_2}- p_{v_2|u_2} \\ 
 p_{v_1|u_3}, & p_{v_2|u_3}, & p_{v_3|u_3}  = 1 -  p_{v_1|u_3}- p_{v_2|u_3}  \end{array} \right)$$
As we will see below, $N \geq 3$ is required because some crucial properties 
of  $\Pi_{v|u} $ do not show up for $N=2$. 
Let us also define a ``return"
probability matrix $\Pi_{u|v}$, by exchanging the roles of the initial and final contexts. 
The matrix $\Pi_{u|v}$ is stochastic like $\Pi_{v|u}$, but these two matrices 
are {\it a priori} unrelated, whereas it is known that in standard QM, they are transpose 
of each other. 
\\

%\vskip 3 mm
%\subsection*{Extra-contextuality of modalities.}
\noindent \textbf{Extra-contextuality of modalities.}
\vskip 2mm

An essential ingredient for determining the mathematical structure of $\Pi_{v|u} $  is provided by a physical constraint 
on the probability $p_{v_j|u_i}$. 
This  is found in Axiom~1~(iii) where it was  stated that ``Modalities cannot show up independently of a context, but 
the same modality may appear in different contexts, with the same conditions of repeatability and certainty".
This means in particular that if $p_{v_j|u_i}= 1$, then $v_j \equiv u_i$, i.e. $v_j $ and $u_i$ represent the same modality, 
within two different contexts. This claim may seem surprising since the measured physical quantities in the two 
contexts can be quite different (see example in Methods); but what matters here is that the certainty and reproducibility
are transmitted from one context to the other, hence the idea that the modality is conserved.  This has also a major 
mathematical consequence, which is that when $p_{v_j|u_i} = 1$, we will require that the {\bf same mathematical object} is associated
with the modality ($v_j $ or $u_i$) in the two contexts. 

More generally, and again in agreement with the physical reality of modalities,  we will require that the probability
$p_{v_j|u_i}$ depends only on the particular modalities $u_i$ and $v_j$ being considered, 
and not on the whole contexts in which they are embedded.
Importantly, to build our formalism, we shall apply this requirement not only to the value of $p_{v_j|u_i}$, 
but also to its mathematical expression; how to do that will be spelled out below. 
This property will be called {\bf ``extra-contextuality" } (see relation with other works in Methods), and it means also that a modality
can be defined  independently of  the $(N-1)$ other modalities  which appear in a given context. 
Such an extra-contextuality is fully compatible with contextual objectivity \cite{ph1,ph2,csm1}~: the latter states that a modality needs a context to be defined, whereas the former  tells that the same modality can show up in several contexts (a simple example is given in Methods, as well
as an interesting link with a proof by John Bell \cite{Bell66}).
\\

%
%\subsection*{Mathematical translation of the Axioms.}
\noindent \textbf{Mathematical translation of the Axioms.}
\vskip 2mm

Summarizing the previous discussions, 
we want to calculate probabilities relating physical events called modalities, occurring 
for a given physical system in a given physical context. Given the physical system, the rules are~:
\begin{itemize}
\item 
{\bf For any given physical context, there are exactly $N$ mutually exclusive modalities.}

As a consequence, the $N^2$ probabilities $( p_{v_j|u_i} )$ connecting the $N$ modalities $\{u_i\}$ (resp.  $\{v_j\}$)
in two different contexts can be arranged  in a $N \times N$ stochastic matrix $\Pi_{v|u}$.
A similar ``return" probability matrix $\Pi_{u|v}$ is defined by exchanging the roles of the initial and final contexts, 
and the set of context transformations  has the structure of a continuous group.
\item 
{\bf If  $p_{v_j|u_i}= 1$, then $v_j $ and $u_i$  are  the same modality, and will be associated 
with the same mathematical object.}

This rule applies within a given context, where one has  $p_{u_j|u_i}=  \delta_{ij}$. If all probabilities $p_{v_j|u_i}$ are either zero or one between two different contexts, one will say that this is the same context, up to re-labelling the modalities. 
\item
{\bf  Extra-contextuality constraint (ECC)~:  the probability $p_{v_j|u_i}$ depends only on the two modalities $u_i$ and $v_j$ being considered,  and not on the whole contexts in which they are embedded. In mathematical terms,  $p_{v_j|u_i}$ depends only on the two mathematical objects associated with the two modalities.}

\end{itemize} 

\noindent These rules have obtained from the Axioms, though not by a fully formal deduction. They may thus  be considered 
as additional principles, deduced from the non-mathematical Axioms, and leading to exploitable mathematical  consequences. 

From there, the main idea of our derivation is the  following: 
we will first write a general parametrization of stochastic 
matrices, which is mathematically and physically neutral, i.e., it is just a rewriting. Nevertheless, 
this parametrization provides a simple criterion for the stochastic matrix to be {\it unistochastic},  
i.e. that its coefficients are the square  moduli of the coefficients of a unitary matrix (see definitions in Methods). 
Then we will translate the extra-contextuality constraint  into an equation, from which we will show  that 
the matrices $\Pi_{v|u}$ and $\Pi_{u|v}$  are unistochastic. 
Finally the usual formalism of quantum mechanics 
(Born's rule, unitary transforms, link between $\Pi_{v|u}$ and $\Pi_{u|v}$) will follow automatically. 
\\

%\subsection*{Mathematical lemmas on stochastic matrices.}
\noindent \textbf{Mathematical lemmas on stochastic matrices.}
\vskip 2 mm

\noindent The theorems below are valid both for $\Pi_{u|v}$ and $\Pi_{v|u}$, so $u$ and $v$ will be omitted 
whenever clarity allows.
\vskip 3 mm

\noindent {\bf Lemma 1:} The elements $p_{j|i} $ of a $N \times N$ stochastic matrix can always  be written under the general form 
\bea
p_{j|i} &=&   \mathbf{Tr}\left( P_i'   \;   R \; P_j''     \;  R \right)
\label{svdth}
\eea
where $\{ P_i'  \}$ and $\{ P_j''  \}$ are two sets of $N$ hermitian projectors of dimension $N \times N$, mutually orthogonal within each set, and where $R$
is a real nonnegative diagonal matrix such that $ \mathbf{Tr}\left( R^2 \right)=N$, and $ \mathbf{Tr}\left(P_i'   \;   R^2 \right)=1$
for all projectors $P_i' $ within $\{ P_i'  \}$.
\vskip 2 mm

\noindent {\it Proof. } Let us first  introduce the
orthogonal ($N \times N$) projectors $P_i$, that are zero everywhere, except for the $i^{th}$ 
term on the diagonal that  is equal to 1; one has obviously  $P_i P_j = P_i \delta_{ij}$.
A useful operation is then to extract the particular probability $p_{v_j|u_i}  $ from  $\Pi_{v|u}$, 
or $p_{u_i | v_j}  $ from  $\Pi_{u|v}$, and
one has the following identities~:
\bea
p_{v_j|u_i} &=&\mathbf{Tr}( P_j \;  \Sigma_{v|u}^\dagger \;   P_i \;  \Sigma_{v|u}) = \mathbf{Tr}(P_i \;  \Sigma_{v|u} \; P_j \;  \Sigma_{v|u}^\dagger)\; \; \; \; \; \;   \label{trace1} \\
p_{u_i|v_j} &=&\mathbf{Tr}(P_i \;  \Sigma_{u|v}^\dagger \; P_j \;  \Sigma_{u|v}) = \mathbf{Tr}(P_j \;  \Sigma_{u|v} \; P_i \;  \Sigma_{u|v}^\dagger)\; \; \;  \; \; \;   \label{trace2}
\eea
where $\mathbf{Tr}$ is the Trace, $^\dagger$ is the Hermitian conjugate, and 
\beq \Sigma_{v|u} = \left[ e^{i \phi_{v_j|u_i} } \sqrt{p_{v_j|u_i}  } \right],  \; 
\Sigma_{u|v} = \left[ e^{i \phi_{u_i|v_j} } \sqrt{p_{u_i|v_j}  } \right] \label{Sigmas} 
\eeq
are $N \times N$ matrices  formed by square roots of the probabilities, and by arbitrary phase factors 
which are introduced here for the sake of generality, and cancel out when calculating  the matrices $\Pi_{v|u}$ and $\Pi_{u|v}$. 

From Eqs. (\ref{trace1}-\ref{Sigmas}) the elements $p_{j|i}$ of a general stochastic matrix $\Pi$ can  be written as
(the subcripts ${u|v}$ or ${v|u}$ are omitted for simplicity):
\beq p_{j|i} = \mathbf{Tr}(P_j \;  \Sigma^\dagger \; P_i  \;  \Sigma ) = \mathbf{Tr}(P_i \;  \Sigma \; P_j \;  \Sigma^\dagger). \label{pji} \eeq

Now, according to the singular values theorem (see Methods), there must exist two unitary matrices $U$ and $V$, 
and a real diagonal matrix $R$, such that 
\beq \Sigma  = U \, R \; V^\dagger, \; \; \; \; \Sigma^\dagger = V \, R \; U^\dagger \label{svdm} \eeq
where the diagonal values of $R$ are the square roots of the (real)  eigenvalues of  
$\Sigma \Sigma^\dagger$, equal to those of  $\Sigma^\dagger  \Sigma$, and are called the singular values of $\Sigma$ (see proof in Methods). 
The value of $ \mathbf{Tr}(R^2)$ is  the sum of the square moduli of all the coefficients of $\Sigma$, and is therefore equal to $N$.

Using Eqs. (\ref{pji}, \ref{svdm}) $p_{j|i}$ can now be written as:
\bea
p_{j|i} &=& \mathbf{Tr}(P_i  \;  U R V^\dagger \; P_j \;  V R U^\dagger) \nonumber \\
&= & \mathbf{Tr}\left( (U^\dagger P_i  \;  U) \;   R \; (V^\dagger  P_j   V)  \;  R \right)\\
&= & \mathbf{Tr}\left( P_i'  \;   R \; P_j''   \;  R \right)
\label{svd}
\eea
where $\{P_i' = U^\dagger P_i  U \}$  and $\{ P_j'' = V^\dagger P_j  V \}$ are 
 two sets of projectors, mutually orthogonal within each set.
 Finally, the normalization condition $\sum_j p_{j|i} =1$   implies that 
\beq
\mathbf{Tr}( P_i \;  \Sigma \; \Sigma^\dagger )  =  \mathbf{Tr}( P_i \;  U R^2 U^\dagger )  =  \mathbf{Tr}( P_i' \;  R^2 )  = 1
\label{norm}
\eeq
hence 
the diagonal elements of $R^2$ 
in the basis associated with the  projectors $\{ P_i '\}$ are all equal to one.  $\square$
 \vskip 2mm

Let us show now
that very different situations occur, depending on the fact that 
 the matrix $R$ is (or is not) equal to the identity matrix $\hat 1$. 
This is related to~: 
 \\
 
  \noindent {\bf Lemma 2:} The  matrix $\Sigma$ is unitary  iff $R=\hat 1$. 
\vskip 2mm

\noindent {\it Proof. } 
Eq. (\ref{svdm}) shows  that  $\Sigma$ is unitary if  $R=\hat 1$, and 
if $\Sigma$ is unitary then $\Sigma^\dagger \Sigma = \hat 1$ and $R^2=\hat 1$, so $R=\hat 1$.  $\square$
\vskip 2mm

An important corollary is that the  matrix $\Pi$ is unistochastic if  $R=\hat 1$. 
The reciprocal is not true, because $\Pi$ being unistochastic does not imply that any matrix 
$\Sigma$  defined by Eq. (\ref{Sigmas}) is unitary  (the phases may be wrong).

An obvious consequence of Lemma 2 is that if $R=\hat 1$ for all possible pairs of contexts, then the  matrix $\Pi$ is unistochastic
for all pairs of context; we will show below that this corresponds to the usual quantum formalism. 
The opposite case is that $R \neq \hat 1$ for some pairs of contexts, but we will show  that 
this contradicts our basic constraint that $p_{v_j|u_i}$ should depend only on the particular modalities $u_i$ and $v_j$ being considered.
First let us establish the following mathematical Lemma:
\vskip 2mm

 \noindent {\bf Lemma 3:}  
 If $R \neq \hat 1$ then ${\cal D} = 0$, where $\cal D $ is the determinant of the unistochastic matrix obtained from the square moduli 
 of the coefficients of  $U$   introduced in Lemma 1. 
\vskip 2mm

\noindent {\it Proof. } 
Let us consider  the normalization conditions  $ \mathbf{Tr}(P_k'   \; R^2 ) = 1$ obtained from Eq. (\ref{norm}),
where the $N$ projectors $P_k'  = U^\dagger P_k U$  correspond to the initial context. 
Since one has also   $ \mathbf{Tr}(P_k'   ) = 1$, the $N$ homogeneous equations 
$ \mathbf{Tr}(P_k' \; ( R^2  - \hat 1) ) = 0$ must be verified by the $N$ diagonal values of $R^2$. 
This set of equations admits a non-trivial solution $R^2 \neq \hat 1$ if its determinant $\cal D$
is equal to zero, and it is easy to check that $\cal D$ is the determinant of the 
unistochastic matrix obtained from $U$. $\square$
\vskip 2mm

Summarizing the previous results, Lemma 1 tells us that for any stochastic matrix $\Pi$,
one can parametrize  the probabilities $p_{j|i}$ by using the diagonal matrix $R$ and  two sets of projectors 
$\{ P_i'  \}$  and $\{ P_j'' \}$.  
Then according to Lemmas 2 and 3,  two situations are possible: 
\vskip 1mm

\noindent - either $R= \hat 1$ for all pairs of contexts, 
and the matrix $\Pi$ is always unistochastic as shown in Lemma 2.
\vskip 1mm

\noindent - or  $R \neq \hat 1$ for some pairs of contexts, 
and a stochastic (but generally not unistochastic)
matrix $\Pi$ is obtained for appropriate projectors, with ${\cal D}=0$  as shown in Lemma~3. 
\\

 %\vskip 2mm
%\subsection*{The Fundamental Theorem}
\noindent \textbf{The Fundamental Theorem}
\vskip 2mm

We are now in a position to use  the extra-contextuality constraint (ECC), which says that  the expression of 
$p_{j|i} =  \mathbf{Tr}( P_i'   \;   R \; P_j''     \;  R )$
should depend only on the particular modalities $u_i$  and $v_j$  being considered, 
and not on the whole contexts in which they are embedded. A first step is the following Lemma:
\vskip 2mm

 \noindent {\bf Lemma 4:}  
Given a N dimensional system, each context must be associated with a set  of $N$ mutually orthogonal projectors, each projector corresponding to one of the $N$ mutually exclusive modalities. 
\vskip 2mm

\noindent {\it Proof. } 
In the case where the initial and final contexts are the same, then  $\Pi = \hat 1$, $\Sigma$ is unitary and diagonal, and $R = \hat 1$. From its definition $V$  can be any unitary matrix, and $U = \Sigma V$, so the two sets of projectors $\{ P_i' \}$ and $\{ P_j'' \}$ are identical, and are associated with the current context. In addition, since Lemma~2 gives $p_{j|i} =   \mathbf{Tr}( P_i'  P_j'  ) = \delta_{ij}$,
each modality $u_k$ ($k=i$ or $j$) must be associated with a projector $P_k'$ of the set $\{ P_k' \}$ corresponding to the current context.  $\square$
%Using Lemma 2 when initial = final $\square$
\vskip 2mm

For $N \geq 3$,  these $N$ projectors may be part of other orthogonal sets, and the corresponding modalities may be part of other contexts. Again for consistency with  the ECC,  we will require that the same projector always corresponds to the same modality. This will extend first  to all contexts containing one (or several) of the initial modalities, giving new projectors and new modalities, and then to the whole space of all $N \times N$ projectors, which will thus be associated to all possible modalities.  This association has to be consistent when the contexts are changed; this will be discussed  in eqs. (\ref{Pi}, \ref{Qi}). 

Let us emphasize that at this point we don't have QM yet; in some sense, we have justified the Hilbert space framework of Gleason's theorem, as the space of $N \times N$ projectors, but we still miss the main hypothesis and the result of the theorem, i.e. Born's law. More precisely, we have justified that 
$P_i'$ and  $P_j''$ depend solely on $u_i$ and $v_j$ in eq. (\ref{svdth}); however, it is still possible that  $R$ depends on the whole contexts $C_u$ and $C_v$  in which $u_i$ and $v_j$ are embedded, and not on these two modalities only. 

So we will use again the ECC to require that not only $P_i'$ and  $P_j''$ but also 
 $R$ depend solely on $u_i$ and $v_j$; 
  more explicitly, this can be written:
\begin{equation}
p_{j|i} =  \mathbf{Tr} ( P_{u_i}'   \;   R_{u_i , v_j} \; P_{v_j}''     \;   R_{u_i , v_j} )
\label{genie}
\end{equation}
where $R_{u_i , v_j} $  depends only on the two specific modalities $u_i, \; v_j$ associated with the projectors  $P_{u_i}',   \;   P_{v_j}'' $, 
and not on the contexts in which they are embedded. 
\vskip 2mm
  
 \noindent {\bf Fundamental Theorem:}   If each modality is bijectively associated 
 with a rank-one projector, and if 
$p_{j|i}$  is given by 
Eq.(\ref{genie}), then  $R_{u_i , v_j}= \hat 1$ for all pairs of contexts. 
  \vskip 2mm

\noindent {\it Proof. }  
Let us assume that $R_{u_i , v_j}  \neq \hat 1$; then one has ${\cal D} = 0$ according to Lemma 3. 
But  one can change the initial context by choosing new projectors  $\{ \tilde P_k '\}$ to replace the  
$\{ P_k '\}$,  keeping $\tilde P_i' = P_i'$ for the modality $i$ of interest, 
whereas the other projectors are different, 
but still mutually orthogonal (this is possible only if $ N \geq 3$). 
Then $ \cal \tilde D$ will generally not be zero any more (see Methods), 
so  the assumption $R_{u_i , v_j}  \neq \hat 1$ is not acceptable.
More generally, the same reasoning is valid for any pair of modalities $u_i$, $v_j$
therefore one has  $R_{u_i , v_j}= \hat 1$ for all pairs of modalities, and also for all pairs of contexts.  $\square$ 
 \vskip 2mm
 
 Intuitively, the theorem says  that the resulting formula 
 $p_{j|i} =  \mathbf{Tr}( P_{u_i}'   \;  P_{v_j}''  )$  provides the unique way   to express
the coefficients $p_{j|i} $ of a stochastic matrix as a function
of the sole modalities $u_i$ and $v_j$, satisfying  the ECC as expressed by  Eq.(\ref{genie}).
We will show now that Born's rule and unitary transforms directly follow from this result. 
 \\
 %\vskip 2mm
 
%\subsection*{Unitary matrices and Born's formula}
\noindent \textbf{Unitary matrices and Born's formula}
\vskip 2mm

From now on we will take $R = \hat 1$ according to the previous Theorem. 
Therefore the matrix $\Sigma_{v|u}= U V^\dagger$ is unitary, but 
one may wonder whether  orthogonal (real) matrices might be enough. 
In order to justify that the full unitary set  is required, we will use Axiom 3, telling that 
the change of contexts corresponds to a continuous group, to require that the set of 
matrices $\Sigma_{v|u}$ is connected in a topological sense, and contains the identity matrix. 
This set must contain permutation matrices,  because they correspond simply to ``relabelling" the modalities, 
i.e. to a trivial change of context.  One has then 
\vskip 2mm
%\\

 \noindent {\bf Lemma 5:}  If  the set of  matrices $\Sigma_{v|u}$, including permutation matrices,
 is connected in a topological sense, and contains the identity matrix, 
then the matrices $\Sigma_{v|u}$ must be complex unitary matrices. 
%\\
\vskip 2mm

\noindent {\it Proof. } 
 The set of  real orthogonal matrices is topologically disconnected 
 in two parts with determinant +1 and -1, 
 whereas permutation matrices 
may have determinant -1, and the identity has determinant +1. 
 On the other hand, all (complex) unitary matrices are connected to the identity, 
 hence the result  (see also refs.\cite{david,aaronson}). $\square$
\vskip 2mm
%\\

We are thus lead to the conclusion that
$\Sigma_{v|u} $ must be a unitary matrix $S_{v|u} $, 
with $S_{v|u} ^\dagger = S_{v|u}^{-1}$.
Then Eqs. (\ref{trace1}) for picking up a particular probability become:
\begin{equation}
p_{v_j|u_i} = \mathbf{Tr}( P_j \,.\,   S_{v|u}^\dagger \,.\,  P_i \,.\, S_{v|u}) =
 \mathbf{Tr}(  P_i \,.\, S_{v|u} \,.\,P_j \,.\,  S_{v|u}^\dagger ).   \nonumber
\label{equ2}
\end{equation}
As said above   $\Pi_{v|u} $ is unistochastic, 
and we can define 
 \bea 
 P_i' =  \; S_{v|u}^\dagger \, . \,  P_i  \, . \,   S_{v|u}\; \; , \; \; \; P_j''=  \; S_{v|u} \, . \,  P_j  \, . \,   S_{v|u}^\dagger  \; \;   \label{projP}.
 \eea
 It is clear that these operators are all Hermitian projectors, i.e.  one has $P^\dagger = P$ and $P^2 = P$ for each of them,
 and also that all sets $\{ P_i' \}$ and $\{ P_j'' \}$
have the same orthogonality properties as  the initial set of projectors $\{ P_i \}$,
i.e. $P_i P_j = P_i \delta_{ij}$. 
One can thus rewrite Eq. (\ref{trace1}) as:
\bea
p_{v_j|u_i} =  \mathbf{Tr}(P_j \;  P_i' ) = \mathbf{Tr}(P_i \;  P_j'' ).  \label{equP} 
\eea
which is just  Born's formula (Eq. \ref{born1}).
Eqs. (\ref{projP}, \ref{equP}) are consistent with our initial requirement 
associating a projector with a modality in  any context, but make clear that this 
association is up to a global unitary transform, related to the choice of a fiducial context. 
In particular, there are two possible choices  for the matrix $\Pi_{v|u} $:
\bea
\text{old context} \{u_i \}&\rightarrow& \text{new context} \{v_j \}  \label{Pi}  \\
P_i' =  \; S_{v|u}^\dagger \, . \,  P_i  \, . \,   S_{v|u} &\rightarrow& P_j \nonumber \\
P_i  &\rightarrow& P_j''=  \; S_{v|u} \, . \,  P_j  \, . \,   S_{v|u}^\dagger \nonumber 
\eea
One can now come back to the matrix $\Pi_{u|v}$, for which the same reasoning is valid, 
and leads to a unitary matrix $S_{u|v}$. By reverting the contexts one has thus:
\bea
\text{old context} \{v_j \}&\rightarrow& \text{new context} \{u_i \}   \label{Qi} \\
Q_j''=  \; S_{u|v}^\dagger \, . \,  P_j  \, . \,   S_{u|v} &\rightarrow& P_i \nonumber \\
P_j &\rightarrow& Q_i' =  \; S_{u|v} \, . \,  P_i  \, . \,   S_{u|v}^\dagger  \nonumber 
\eea
Again, the projectors should be the same for a given modality  in a given context, 
i.e. one should have $P_j''=Q_j''$ (for the same $P_i$ in the other context), 
and $P_i' =Q_i' $ (for the same $P_j$ in the other context). 
This is obtained if $S_{u|v}$ is the inverse of $S_{v|u}$,
leading to a last lemma:
\vskip 2mm

\noindent {\bf Lemma 6:}
 If $P_j''=Q_j''$ and $P_i' =Q_i' $ as defined above, then
$S_{u|v} = S_{v|u}^\dagger= S_{v|u}^{-1} $ (up to global phase factors), and  
 the matrices $\Pi_{u|v} $ and $\Pi_{v|u} $ are related by $\Pi_{u|v} = \Pi_{v|u}^{t}$. 
\vskip 2mm

\noindent  {\it Proof: } Obvious from the relations (\ref{Pi}) and  (\ref{Qi}) $\square$. 
\vskip 2mm
%\\

Then the various points of view represented in the relations (\ref{Pi}, \ref{Qi})
are all consistent and give the same values for the probabilities, because 
each $S_{v|u}$  can be associated to an element of the group of context transformations ${\cal G}$, and 
its inverse is $S_{u|v} = S_{v|u}^{-1} = \, S_{v|u}^\dagger$. 
For the general consistency of the approach including Axiom 3, 
this set of matrices gives a $N \times N$ (projective) representation of the group of context transformations;
this is fully consistent with the well known Wigner theorem \cite{fl}.
This continuous unitary evolution will be essential 
to describe the evolution  of the system (translation in time) \cite{fl}. 
Since we have now reached the starting point of most QM textbooks \cite{cct},   it should be clear that 
the standard structure of QM can be obtained from this construction. In particular, one can 
associate the $N$ orthogonal projectors $\{ P_i \}$ to the $N$  orthonormal vectors
which  are eigenstates of these projectors up to a phase factor, i.e., to  rays in the Hilbert space.
Similarly, the expected probability law  for the measurement results $\{a_i \}$ will be obtained 
by writing any physical quantity $A$ as an operator $\hat A = \sum a_i P_i$, 
this is  the usual spectral theorem.  The tensor product structure for composite systems 
can also be introduced in the usual way  (see also the last section of Methods).  
\\

%\section*{Discussion.}
%\subsection*{Discussion.}
\noindent \textbf{Discussion.}
\vskip 2mm

An interesting  outcome of our derivation is that the usual Hilbert space structure (for $ N \times N$ matrices)
shows up, without any initial assumption of a superposition principle, interference effect, or wave function \cite{r1,r2,r3z,r3,r4,rr4,r5}. 
This structure comes directly from requirements on probabilities, implying that  the matrices
$\Pi_{u|v}$ and $\Pi_{v|u}$  belong to the unistochastic subset of stochastic matrices. 
This appears as the mathematical consequence of the joint physical requirements of  
contextuality of the theory (contexts are needed to define modalities),  
quantization of modalities (making probabilities necessary), 
and  extra-contextuality of  modalities 
(probabilities depend on modalities, that may belong to different contexts). 

Extra-contextuality  is also a crucial hypothesis
for Gleason's theorem \cite{gleason,helena}, which is deeply related to our derivation; however, the reasonings proceed
in quite different ways.  The Hilbert space structure is an assumption in Gleason's theorem, whereas
in our case it appears more heuristically. 
Rather than reconstructing ``from scratch" the Trace formula,  as done by Gleason, 
we introduce it as a general parametrization of stochastic matrices; this avoids 
the heavy machinery of Gleason's theorem \cite{gleason,helena}, in particular the demonstration of continuity (see Methods).
Then we use extra-contextuality 
to restrict acceptable matrices to unistochastic ones, ending up again with Born's formula in finite dimension. 
Using explicitly Gleason's theorem is also possible \cite{csm2}, and 
%with  the advantage of  lifting 
%allows one to lift 
removes the restriction on a finite $N$  (see Methods). 

We emphasize that we do not need
any additional ``measurement postulate", since measurement is already included
in Axiom 1, i.e. in the very definition of a modality \cite{csm1,csm3,ph1,ph2}. 
Quantum superposition are here as usual, but they are not spooky 
``dead-and-alive" concepts: they are rather the manifestation of a modality (i.e., a certainty) in
another context. Entanglement  is also present as linear superpositions of tensor product states, 
corresponding to modalities in a ``joint" context, and the specific case of 
 two-particle Bell-EPR experiment is discussed in ref.\cite{csm3}. 
Since a modality requires both a context and a system, 
it embeds non-local features corresponding to quantum non-locality, 
but it is fully compatible with relativistic causality  \cite{csm1,csm3}, and 
operationally agrees with no-signaling, just like QM does. 
From a foundational point of view, our approach  also provides a clear  distinction between
the modality, which is a real physical phenomenon, or a physical event in the sense of 
probability theory, and the projector, 
which is a mathematical tool for calculating non-classical probabilities. 

To conclude, let us emphasize that we discussed a very idealized version of QM, 
based on pure states and orthogonal measurements. Nevertheless, this idealized
version does provide the basic quantum framework,  and
connects the experimental definition of a physical quantity and the measurement results 
in a consistent way, both physically and philosophically  \cite{csm1}.
Adding more refined tools such as density matrices, imperfect measurements, POVM, open systems, decoherence,
is of great practical interest and use, but this will not ``soften"  
the basic ontology of the theory, as it is presented here. 
The present work, deeply rooted in ontology, is thus  complementary to many recent related proposals
 \cite{zurek,r1,r2,r3z,r3,r4,rr4,r5,r6,r7,r8,rr8,r9}.

\section*{Methods}

{\centerline {\it Stochastic matrices. }}
\vskip 2mm
%\bibitem{bist3} 
A {\it stochastic} matrix has real positive coefficients, with all lines summing up to 1.
{\it Bistochastic} matrices are stochastic ones, with both  lines and columns summing to~1.
{\it Orthostochastic} and {\it unistochastic} matrices are
obtained by taking the square moduli of the coefficients of respectively 
an orthogonal or a  unitary matrix \cite{unis}. 
For $N=2$, all bistochastic matrices  are also ortho- and uni-stochastic, and 
for $N \geq 3$,  the set of unistochastic matrices is larger than the orthostochastic set, 
but smaller than the bistochastic  set.  
For instance, the simple matrix
$$ {\frac{1}{2}
{%\tiny 
\left(  \begin{array}{lll} 
1&1&0  \\ 
0&1&1 \\ 
1&0&1  \end{array} \right)}}$$
is a well known example of a bistochastic matrix, which is neither  orthostochastic, nor unistochastic; 
therefore it is not an acceptable (quantum) probability matrix $\Pi_{v|u} $. 
\vskip 4mm

%\subsubsection*{Singular values theorem and the invariance of $R$.  }
%\vskip -3mm
{\centerline {\it Singular values theorem and the invariance of $R$.  }}
\vskip 2mm
%\bibitem{dem-svd}  
To obtain the singular values decomposition, 
diagonalize the Hermitian matrix $\Sigma^\dagger \Sigma$, get the real diagonal matrix $R$ and 
the unitary matrix $V$ so that  $V^\dagger \Sigma^\dagger \Sigma V = R^2$.
Then define another unitary matrix $U$ such that $U R =  \Sigma V$, and 
$R U^\dagger =  V^\dagger \Sigma^\dagger  $,   so that $ U^\dagger \Sigma \Sigma^\dagger U = R^2 $.
One gets  thus the decomposition $\Sigma = U R V^\dagger$ as expected. 
%\\

%\bibitem{det} 
In the demonstration of the fundamental theorem, we note that one might restrict the new projectors  $\{ \tilde P_k '\}$ to be such that ${\cal \tilde D} = 0$. Then $R$ can be different of $\hat 1$, but it has still to be modified to some $\tilde R$ to fulfill the normalization conditions with the $\{ \tilde P_k '\}$. Since the hypotheses is  that $R$ should be constant, this case is excluded also. 

One may also wonder what would happen if no phase factors were included in the definition of $\Sigma$. Then the Lemmas are still valid, but $\Sigma$ cannot be unitary, and not even orthogonal. Then according to Lemma 2, $R$ cannot be the identity, and therefore the extra-contextuality constraint cannot be satisfied. 
\vskip 4mm

{\centerline {\it Extra-contextuality and Gleason's theorem. }}
\vskip 2mm
%\bibitem{nonc} 
Extra-contextuality is not a new concept, but it is a new name given to a known concept, 
called non-contextuality in articles dealing with Gleason's theorem \cite{helena}, 
or  ``measurement non-contextuality"  in  ref.\cite{r8}. Extra-contextuality  is {\bf not} the contrary of contextuality,
and it avoids confusion arising when using ``non-contextuality". In particular, contexts are needed to define modalities, 
and modalities are extra-contextual,  without any contradiction with refs.\cite{ph1,ph2,csm1,csm2} or with  the Kochen-Specker theorem. 
%\bibitem{spins} 
As a simple example of extra-contextuality, consider a system of two spin 1/2 particles, and define $\vec S = \vec S_2  + \vec S_2$. Using standard notations
for coupled and uncoupled basis,  the $|m_1=1/2, m_2=1/2 \rangle$ modality in the context $\{S_{z1},S_{z2} \}$ is the same 
as the $|S=1, m_S=1 \rangle$ modality in the context $\{\vec S^2, S_{z} \}$, though other modalities in the same  two contexts are different.  

Demonstrating continuity of the probability formula is an essential step of Gleason's theorem. 
In our derivation continuity appears formally in Axiom 3, and is embedded in the matrix formalism
that we are using. It is used  in the Fundamental theorem to to build a new set of projectors 
$\{ \tilde P_k '\}$, keeping one of them constant, and  in Lemma 5 to get complex unitary matrices. 
So it does play a role, but does not have to be demonstrated. 
The explicit use of Gleason's theorem for allowing  the dimension $N$ to be infinite is spelled out in ref.\cite{csm2}. 
This requires  to introduce an Axiom 4 associating explicitly modalities and projectors in an Hilbert space; 
such an Axiom is not formally required in the present heuristic derivation, but it provides a useful ``back-up". 

It is interesting to note that John Bell demonstrated explicitly in  ref.\cite{Bell66} (Section V) that if extra-contextuality is accepted  as it is done in the hypothesis of Gleason's theorem, then the impossibility of hidden variables  (HV) automatically follows.
More technically, Bell showed that there is no dispersion-free state, then he wrote at the end of his proof~: 
{\it It was tacitly assumed that measurement of an observable must yield the same value independently of what other (commuting) measurements
may be made simultaneously. Thus as well as $P(\Phi_3)$ say (projector on vector $\Phi_3$), one might measure  {\it either $P(\Phi_2)$ or $P(\Psi_2)$}, where $\Phi_2$ and $\Psi_2$ are orthogonal to $\Phi_3$ but not to one another. These different possibilities require different experiment arrangements; there is no {\it a priori} reason to believe that the results for $P(\Phi_3)$ should be the same.} So Bell rejected the ``tacit assumption'', and therefore also the conclusion that there is no dispersion-free states. 
But extra-contextuality, seen as a consequence of  the reality of modalities,  may provide the missing ``a priori  reason" to accept the assumption, and thus also Bell's proof that there is no dispersion-free state.

Obviously Bell's goal was very different from ours, since he was investigating the possibility of HV, whereas we want to recover the quantum formalism. Nevertheless, we conclude that if we accept extra-contextuality, 
then the quantum formalism follows (from the present article), and consistently with this result, HV are excluded (from  the 
Bell-Gleason argument of ref.\cite{Bell66}). 
\vskip 4mm

{\centerline {\it Relations with textbook quantum mechanics. }}
\vskip 2mm
In this section we outline various issues relating our approach to standard QM.
First, we considered only pure states and orthogonal (projective) measurements. 
This fits with  the usual view that mixed states (density matrices)  and non-orthogonal measurements (POVM) 
correspond to more classical aspects of probabilities, and can be introduced at a later stage. This is possible because 
in each context, a classical probability distribution can be built upon the $N$ mutually exclusive modalities. 

In our approach entanglement appears naturally in the following way: let us consider two systems 1 and 2 with $N_1$ and 
$N_2$ mutually exclusive modalities. If both systems are considered together, but each one in  its own context, there are clearly $N = N_1 \times N_2$
mutually exclusive modalities. But from Axiom 2, the value of $N$ does not depend on the context, so the global system must be described 
by $N \times N$ projectors and unitary matrices. Many of these projectors cannot be split into projectors acting separately on system 1 or system 2, 
and are associated to entangled states. 

In this article the postulate  on time evolution is not  spelled out, but it enters in the same framework, 
by including translations in time in the group ${\cal G}$. For instance,  if  ${\cal G}$ is the Galileo group, 
standard non-relativistic QM can be recovered, including Schr\"odinger's equation \cite{fl}. 
Also, we did not discuss the known connection between the physical
quantities and the infinitesimal generators of ${\cal G}$, or the role of ``projective" representations.
Finally, we considered only non-relativistic quantum mechanics, and therefore ``type I" von Neumann algebra, see e.g. ref.\cite{type1}.
\\

%\section*{Acknowledgements}
\noindent \textbf{Acknowledgements}
\vskip 2mm

The authors thank Fran\c cois Dubois, Cyril Branciard  and Anthony Leverrier for many discussions,
and Nayla Farouki for continuous support. 
P.~G.  also  thanks  Ad\`an Cabello and  Alastair Abbott for useful comments and discussions.

\end{document}